# Production of dileptons in ultra-peripheral heavy ion collisions with two-photon processes

Gongming Yu[1], Xinyu Li[2], Xinghang Zhang[2], Zhongxia Zhao[3], Haitao Yang[3]

[1]College of Physics and Technology, Kunming University, Kunming 650214, China
[2]Fundamental Science on Nuclear Safety and Simulation Technology Laboratory, Harbin Engineering University, Harbin 150000, China
[3]College of Science, Zhaotong University, Zhaotong 657000, China

We study the photoproduction process of dileptons in heavy ion collision at Relativistic Heavy Ion Collider (RHIC) and Large Hadron Collider (LHC) energys. The equivalent photon approximation, which equates the electromagnetic field of high-energy charged particles to the virtual photon flux, is used to calculate the processes of dileptons production. The numerical results demonstrate that the experimental study of dileptons in ultra-peripheral collisions is feasible at RHIC and LHC energies.

Key words: Ultraperipheral collisions, Dilepton, Photoproduction process, Equivalent photon approximation

**PACS:** 12.39.St, 13.85.Dz, 25.75.Cj

## I. INTRODUCTION

In the ultra-peripheral collision process of two heavy ions, the vertical distance between the centers of the two heavy ions in the direction of motion is much greater than the sum of their radii. At this distance, there is almost no strong interaction between the two nucleus, and the interaction between the two heavy ions depends on the electromagnetic fields radiated by the two nucleus [1,2]. High-energy heavy ions will emit a strong electromagnetic field around them, which is the main source of virtual photons in photon-induced reactions [2,3]. Therefore, ultra-peripheral collision processes are an important platform for studying photon-induced production, and recent experiments at RHIC and LHC have made many research in this field [4,5].

In the recent years, photoproductions interactions were carried out through lepton beams, where high-energy leptons interacted with target nuclei through the electromagnetic field which is radiated into their surroundings [4-9]. However, compared to lepton beams, the energy of the electromagnetic field emitted by high-energy protons or heavy ions is much higher [10-13]. This makes the ultra-peripheral collision process of high-energy heavy ions as an important platform for studying photon-induced processes. In the photoproduction process, there are two types of reactions: photon-heavy ion interactions and photon-photon interactions. The photon-heavy ion reactions refers to the interaction between the high-energy electromagnetic field radiated by one particle and another particle, which can be regarded as the interaction between the photon and the constituent particle within the quark-parton model. As for photon-photon interactions, it refers to the interaction between particles through the high-energy electromagnetic field respectively radiated by two charged particles, which can be considered as the interaction between two photons [2,3,14]. Therefore, it can be concluded that strong interaction plays a major role in the photon-hadron reaction process, and the final state particles produced are more complex than in the case of photon-photon process. The photon-photon interaction has more cleaner background in compared to photon-hadron processes [15].

Due to the complexity in calculating virtual photons, which leads to a strong correlation between the process of radiation and absorption between virtual photons and particles, it has been found that the excitation process involving transition matrix elements when a nucleus electromagnetic field causes non-elastic transitions in another nucleus is similar to that of real photons. This provides a basis for using photon spectroscopy functions instead of the interaction. Therefore, when studying the photoproduction process, we often use the equivalent photon spectrum approximation method for processing. This method was originally proposed by Fermi in 1924. Fermi suggested an equivalent photon approximation method during the electrical bombardment of atoms with fast alpha particles, treating charged particles as flux of virtual photons. Later, this method was extended to relativistic particles by Weizsacker and Williams, so it is also called the Weizsacker-Williams method [16-19]. The Fast-moving charged particles have radial electric field vectors pointing outward and magnetic fields surrounding them. The field at points a certain distance from the particle trajectory is similar to that of a real photon. Therefore, in the equivalent photon spectrum approximation method, Fermi used an equivalent photon flux to replace the electromagnetic field of fast-moving charged particles. Through this method, the virtuality that runs through the photon radiation process and photon absorption process is set to zero. This can eliminate its impact on the entire experiment and simplify the calculation process. [20-22] At the same time, photon radiation and absorption processes can be calculated separately, making it easier to factorize in subsequent calculations, greatly simplifying the computation. Therefore, this method is often used to calculate the photon-photon process in the ultra-peripheral collision process of relativistic heavy ions. In the present work, we investigatethe large-$p_T$ dilepton production in two-photon interaction processes with the semi-coherent approximation, that a non-coherent photon with large transverse momentum is radiated from a nucleus and the coherent photon with small transverse momentum is radiated from a proton of another nucleus in the relativistic heavy ion collisions.

This paper is organized as follows: In section II, we present the semi-coherently two-photon production process for dileptions at RHIC and LHC energies. The numerical results for dileptions production are plotted in Sec.III. Finally, the conclusion is given in Sec.IV.

## II. GENERAL FORMALISM

It is known that the differential cross section for ultraperipheral collisions of heavy ions is made up of two parts. One is the production of the photon spectrum of each nucleus [23,24], and another is the production of dileptons in two photon collisions. In the equivalent photon approximation, the electromagnetic field of the fast-moving heavy ion is replaced by an equivalent real photon spectral function [25-27]. The cross section for two photon interactions is given by

$$\frac{d\sigma_{\gamma\gamma\to l^+l^-}(M)}{dM} = \hat{\sigma}_{\gamma\gamma\to l^+l^-}(M) = \frac{4\pi\alpha^2 e_f^2}{M^2}\left[2\left(1+y-\frac{1}{2}y^2\right)\ln\left(\frac{1}{\sqrt{y_s}}+\sqrt{\frac{1}{y_s}-1}\right)\right], \quad (1)$$

here, the expression of the variable $y_s$ is $y_s = 4M_{l^+l^-}^2/M^2$, $M$ is the invariant mass of the dilepton, $M_{l^+l^-} = 2M_l$. The masses of the three leptons are $M_e = 0.5109 MeV, M_\mu = 105.66 MeV, M_\tau = 1776.86 \pm 0.12 MeV$.

For the two-photon process in the ultraperipheral collisions, the virtual photon can be treated as real photon since he electromagnetic field equivalent of the heavy ion moving at high speed. In the semi-coherent approximation, the momentum for photons are $q_1=(\omega_1, \boldsymbol{q}_{1T}, q_{1z})$ and $q_2=(\omega_2, \boldsymbol{q}_{2T}, q_{2z})$, the total transverse momentum of bound ($l^+l^-$) state is $\boldsymbol{p}_T = \boldsymbol{q}_{1T} + \boldsymbol{q}_{2T} \approx \boldsymbol{q}_{1T}$, where $\boldsymbol{q}_{iT}$ is the transverse momentum of the $i$-th photon. The differential cross section for nucleus-nucleus collisions can be obtained as

$$\begin{aligned}
d\sigma_{l^+l^-} &= \hat{\sigma}_{\gamma\gamma\to l^+l^-}(M) dN_1(\omega_1, q_1^2) dN_2(\omega_2, q_2^2) \\
&= d\omega_1 d\omega_2 \hat{\sigma}_{\gamma\gamma\to l^+l^-}(M) \frac{dN_1(\omega_1, q_1^2)}{d\omega_1} \frac{dN_2(\omega_2, q_2^2)}{d\omega_2} \\
&= \frac{\alpha^2 Z_1^2 Z_2^2}{\pi^2} \int \frac{d\omega_1}{\omega_1} \frac{d\omega_2}{\omega_2} \hat{\sigma}_{\gamma\gamma\to l^+l^-}(M) \\
&\times \int dq_{1T}^2 q_{1T}^2 \frac{\left[F_N\left(q_{1T}^2+\frac{\omega_1^2}{\gamma^2}\right)\right]^2}{\left(q_{1T}^2+\frac{\omega_1^2}{\gamma^2}\right)^2} \\
&\times \int dq_{2T}^2 q_{2T}^2 \frac{\left[F_N\left(q_{2T}^2+\frac{\omega_2^2}{\gamma^2}\right)\right]^2}{\left(q_{2T}^2+\frac{\omega_2^2}{\gamma^2}\right)^2},
\end{aligned} \quad (2)$$

Then, we can get

$$\frac{d\sigma_{AB \to Al^+l^-B}}{dMdp_T^2dy} = \frac{\alpha^2 Z_1^2 Z_2^2}{\pi^2} \frac{2P_T^2}{M} \hat{\sigma}_{\gamma\gamma \to l^+l^-}(M) \frac{\left[F_N\left(P_T^2 + \frac{\omega_1^2}{\gamma^2}\right)\right]^2}{\left(P_T^2 + \frac{\omega_1^2}{\gamma^2}\right)^2}$$

$$\times \int dq_{2T}^2 q_{2T}^2 \frac{\left[F_N\left(q_{2T}^2 + \frac{\omega_2^2}{\gamma^2}\right)\right]^2}{\left(q_{2T}^2 + \frac{\omega_2^2}{\gamma^2}\right)^2}, \quad (3)$$

where the energies for the photons emitted from the nucleus are $\omega_{1,2} = \frac{W}{2}\exp(\pm y)$, with $W^2 = 4\omega_1\omega_2$, and the transformations $d\omega_1 d\omega_2 = (W/2)dWdy$ can be performed.

The equivalent photon spectra from the nuclei with $Z$ times the electric charge or proton moving with a relativistic factor $\gamma \gg 1$ can be written as[21,28-30]

$$\frac{dN_N(\omega, q^2)}{d\omega} = \frac{Z^2\alpha}{\pi\omega} \int d^2q_T \frac{q_T^2}{\left(q_T^2 + \omega^2/\gamma^2\right)} F_N^2(q^2),$$

$$\frac{dN_p(\omega, q^2)}{d\omega} = \frac{Z^2\alpha}{\pi\omega} \int d^2q_T \frac{q_T^2}{\left(q_T^2 + \omega^2/\gamma^2\right)} F_p^2(q^2),$$

where $q^2 = (q_T^2 + \omega^2/\gamma^2)^2$ is the 4-momentum transfer of nuclei or proton, the nuclear form factor of nuclei and proton can be written as[21,28-30]

$$F_N(q^2) = \frac{\Lambda^2}{\Lambda^2 + q^2}, \quad \text{where} \quad \Lambda^2 = \frac{0.164\text{GeV}^2}{A^{2/3}},$$

$$F_p(q^2) = \frac{1}{\left(1 + \frac{q^2}{0.71\text{GeV}^2}\right)^2},$$

where $\Lambda = 0.091$GeV for $^{197}$Au, $\Lambda = 0.088$GeV for $^{208}$Pb, and $\Lambda = 0.065$GeV for $^{238}$U.

Similarly, the differential cross section for proton-nucleus and proton-proton collisions collisions the can be obtained as

$$\frac{d\sigma_{pA \to pl^+l^-A}}{dMdp_T^2dy} = \frac{\alpha^2 Z^2}{\pi^2} \frac{2P_T^2}{M} \hat{\sigma}_{\gamma\gamma \to l^+l^-}(M) \times \left\{ \frac{\left[F_p\left(P_T^2 + \frac{\omega_1^2}{\gamma^2}\right)\right]^2}{\left(P_T^2 + \frac{\omega_1^2}{\gamma^2}\right)^2} \int dq_{2T}^2 q_{2T}^2 \frac{\left[F_N\left(q_{2T}^2 + \frac{\omega_2^2}{\gamma^2}\right)\right]^2}{\left(q_{2T}^2 + \frac{\omega_2^2}{\gamma^2}\right)^2} \right.$$

$$\left. + \frac{\left[F_N\left(P_T^2 + \frac{\omega_1^2}{\gamma^2}\right)\right]^2}{\left(P_T^2 + \frac{\omega_1^2}{\gamma^2}\right)^2} \int dq_{2T}^2 q_{2T}^2 \frac{\left[F_p\left(q_{2T}^2 + \frac{\omega_2^2}{\gamma^2}\right)\right]^2}{\left(q_{2T}^2 + \frac{\omega_2^2}{\gamma^2}\right)^2} \right\}, \quad (4)$$

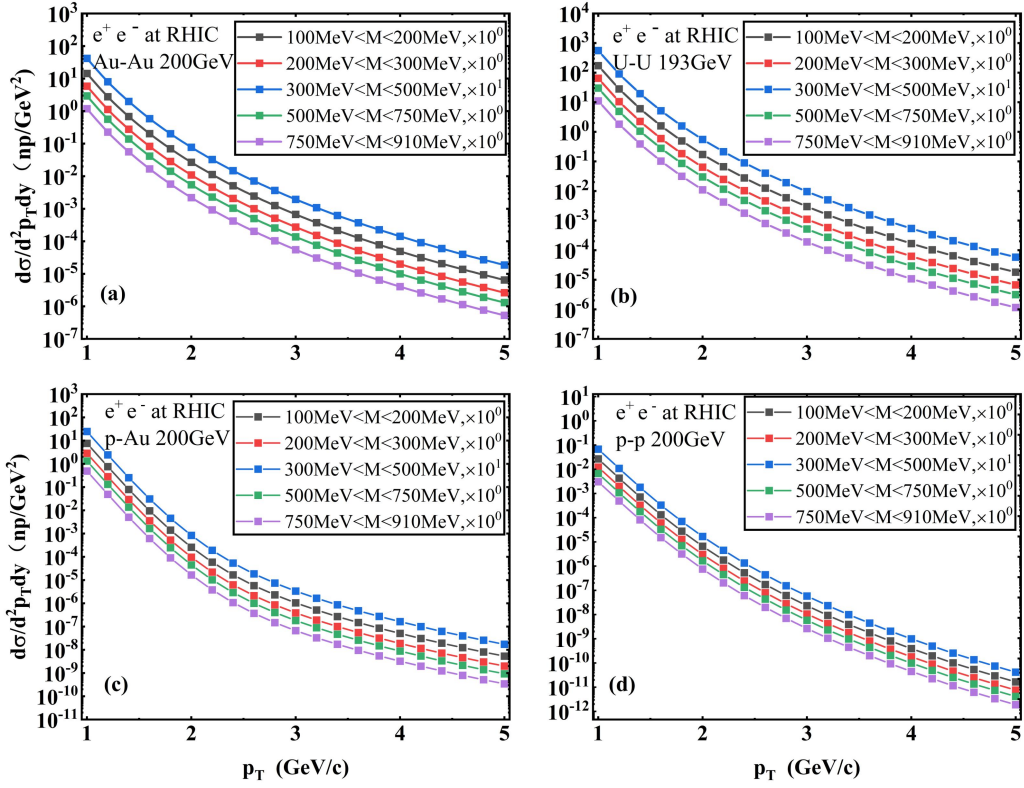

Figure 1 The cross sections for e⁺e⁻ production from the semicoherent two-photon interaction at RHIC.

$$\frac{d\sigma_{pp\to pl^+l^-p}}{dMdp_T^2dy} = \frac{\alpha^2}{\pi^2}\frac{2P_T^2}{M}\hat{\sigma}_{\gamma\gamma\to l^+l^-}(M)\times\frac{\left[F_p\left(P_T^2+\frac{\omega_1^2}{\gamma^2}\right)\right]^2}{\left(P_T^2+\frac{\omega_1^2}{\gamma^2}\right)^2}\int dq_{2T}^2 q_T^2\frac{\left[F_p\left(q_{2T}^2+\frac{\omega_2^2}{\gamma^2}\right)\right]^2}{\left(q_T^2+\frac{\omega_2^2}{\gamma^2}\right)^2}, \quad (5)$$

Here $M$ is the invariant mass of dilepton.

III. NUMERICAL RESULTS

Figure 1 and 2 is the cross-sectional distribution of the electron pair produced in the two-photon process at the RHIC energy region. The black line indicates that the constant mass interval of the electron pair is $100 MeV < M < 200 MeV$, the red line is $200 MeV < M < 300 MeV$, the blue line is $300 MeV < M < 500 MeV$, the green line is $500 MeV < M < 750 MeV$, and the purple line is $750 MeV < M < 910 MeV$. For the photoproduction process of electron pairs, it can be seen from Figures 1 and 2 that the contribution of the double-photon process of electron pairs shows an exponentially decreasing trend with respect to the transverse momentum $p_T$ in the calculation region after taking logarithmic conversion. The invariant mass of the electron pair increases, the production cross-section for the photoproduction process also shows a decreasing trend.

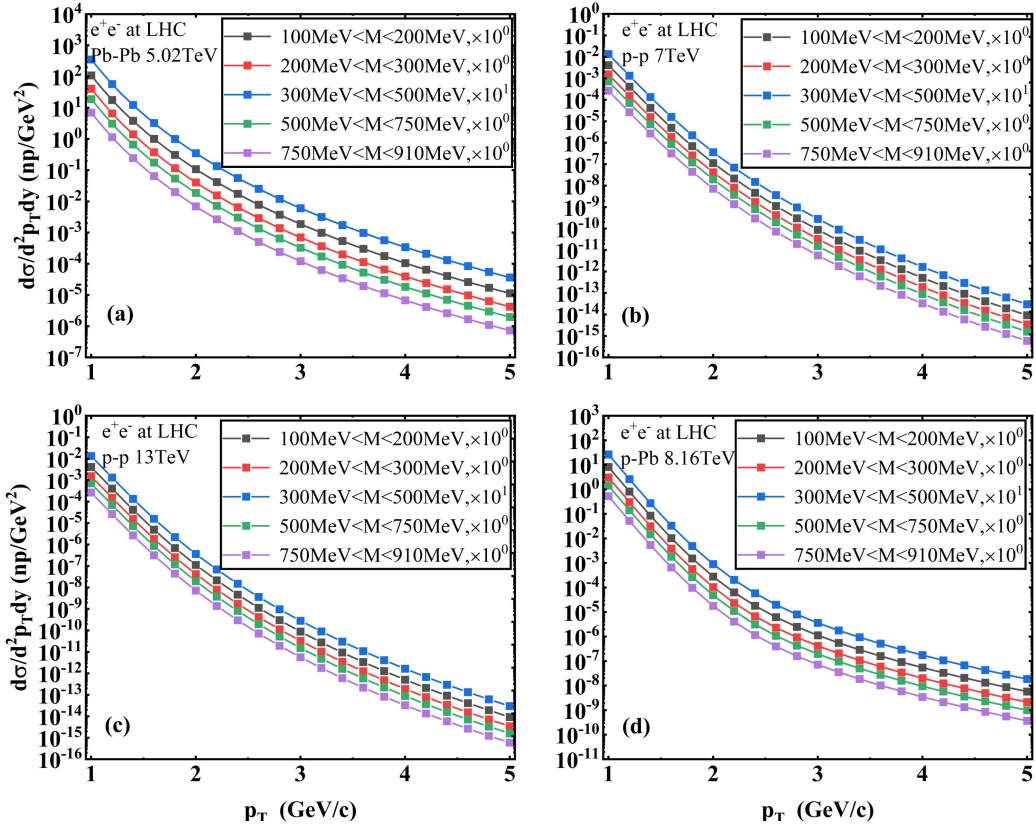

Figure 2 The cross sections for e⁺e⁻ production from the semicoherent two-photon interaction at LHC.

By analyzing the cross-section distributions of p-Au and p-p at the same center-of-mass energy, it can be found that the differential cross-section for p-Au is greater than that for p-p. This is because in p-Au collisions, there are more particles and higher charge density in Au nuclei, in compared with p-p collisions. When the charge density is higher, the electromagnetic field strength around it would increase, that makes it more likely to undergo electromagnetic interaction, i.e., the probability of the double-photon process occurring is increased. This leads an increase in the cross-section for the photoproduction process of dileptons. In our calculation, the charge distribution is used to approximate the shape factor. This leads a rapid decrease in the production cross-section of dileptons in the $\gamma\gamma \to e^+e^-$ process as the transverse momentum of the electron pair increases. Similarly, the curves in the range of p-p collisions at center-of-mass energies of 200 MeV, 7 TeV, and 13 TeV, it can be seen that the three sets of data have almost no difference. It is speculated that this is due to the very small value of the rest mass of electrons (0.5109 MeV), resulting in a negligible production cross-section for dileptons in double-photon processes. Even with an increase in the center-of-mass energy, the impact on the production cross-section for dileptons is extremely weak and falls within the allowable calculation error, which leads to the same data. In contrast, increasing the charge density of colliding particles has a more pronounced effect on the production cross-section for electron pairs.

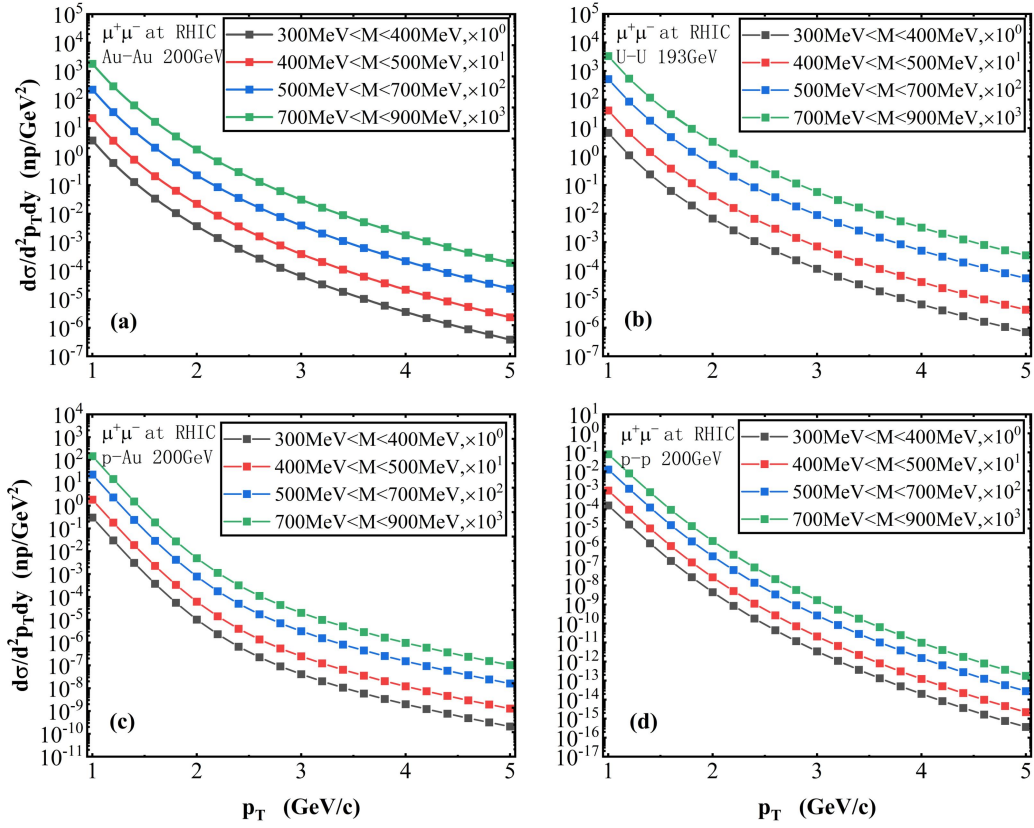

Figure 3 The cross sections for μ⁺μ⁻ production from the semicoherent two-photon interaction at RHIC.

Figure 3 and Figure 4 are the distribution of the production cross-section for muon pairs in double-photon processes at RHIC and LHC energies, respectively. In the figures, the black line represents the invariant mass range of muon pairs between $300 MeV < M < 400 MeV$, the red line is $400 MeV < M < 500 MeV$, the blue line is $500 MeV < M < 700 MeV$, and the green line is $700 MeV < M < 900 MeV$. From Figures 3 and 4, it can be seen that the contribution of double-photon processes to the production of muon pairs for large transverse momentum $p_T$ is basically the similar as the photo-production of electron-positron pairs, but the decreasing trend of muon pairs is slightly larger than that of electron pairs due to the higher rest mass of muons (105.66 MeV) than electrons. Under the same conditions, particles with higher mass usually have higher electromagnetic responses, meaning that they absorb more energy and thus have higher production cross-sections.

Figures 5 and 6 show the distribution of cross-sections for tau pairs produced in photon-photon interactions at RHIC and LHC energy regions, respectively. The black line corresponds to the invariant mass interval of $4 GeV < M < 5 GeV$, the red line corresponds to $5 GeV < M < 6 GeV$, the blue line is to $6 GeV < M < 7 GeV$, and the green line is to $7 GeV < M < 9 GeV$.

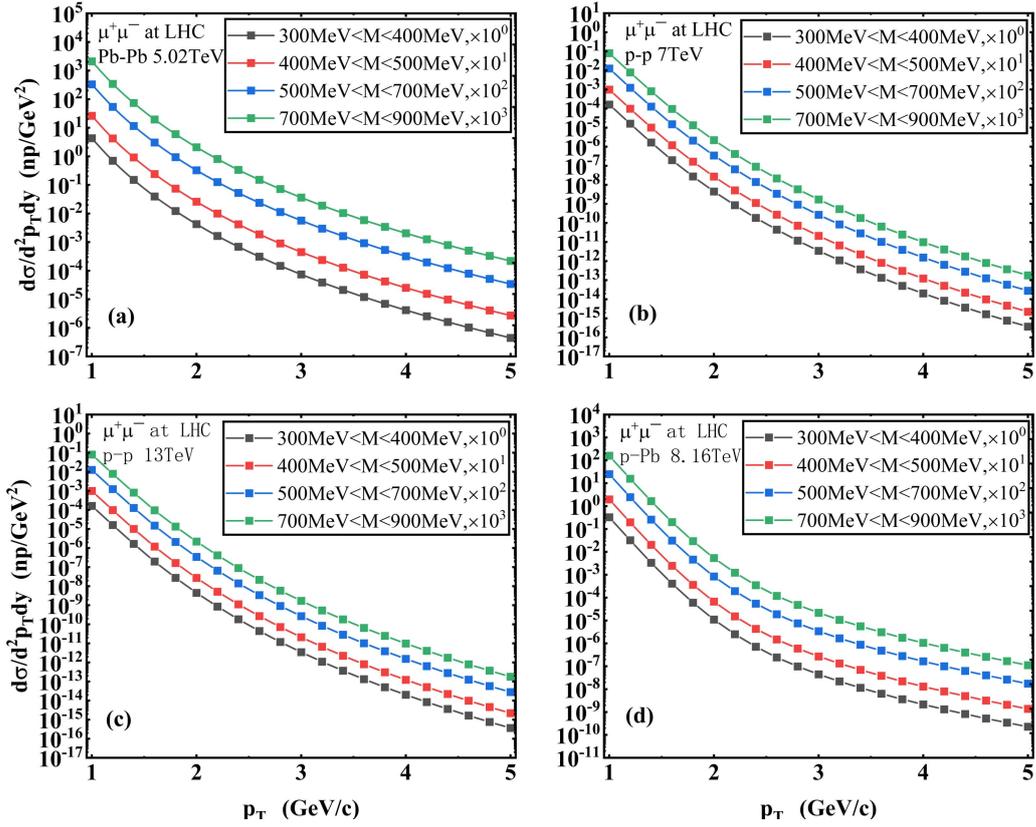

Figure 4 The cross sections for μ⁺μ⁻ production from the semicoherent two-photon interaction at LHC.

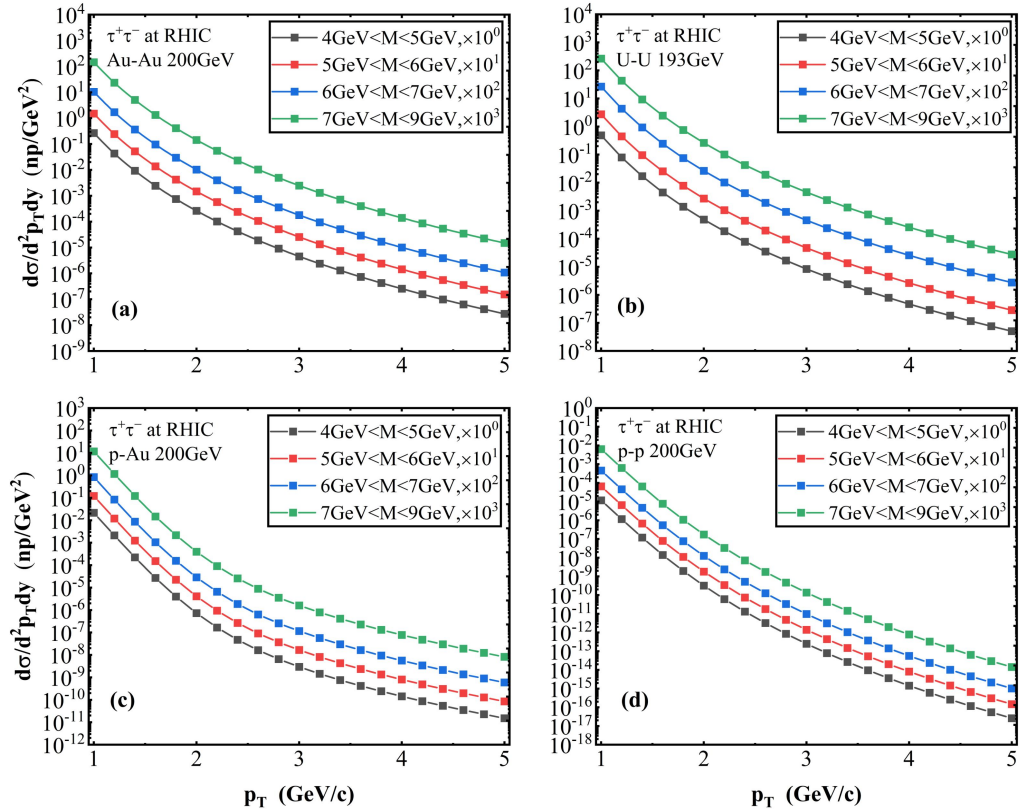

Figure 5 The cross sections for τ⁺τ⁻ production from the semicoherent two-photon interaction at RHIC.

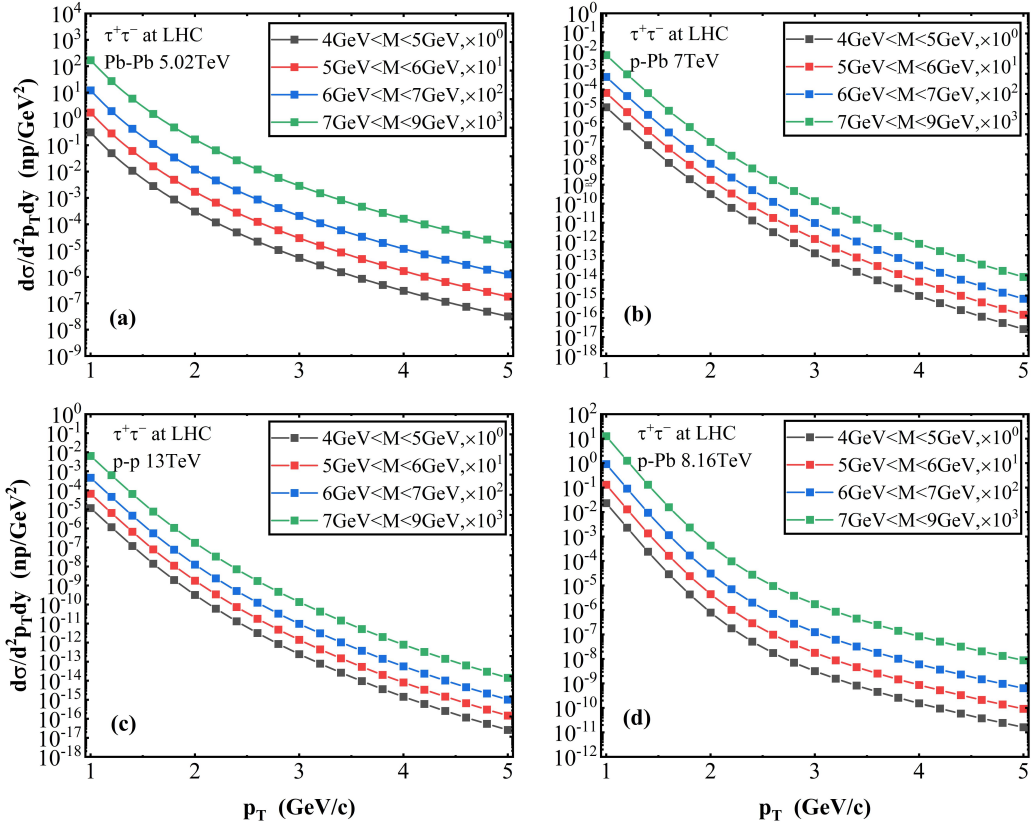

Figure 6 The cross sections for $\tau^+\tau^-$ production from the semicoherent two-photon interaction at LHC.

## IV. CONCLUSION

In summary, have investigated the photoproduction of dileptons with large transverse momentum in ultraperipheral collisions with semi-coherently two-photon production process. For the photoproduction process of double leptons, the relationship between its production cross-section and the transverse momentum $p_T$ shows an exponential decrease. The production cross-section decreases as the invariant mass of the dileptons increases. This is because production of heavier dilepton need higher total energy and momentum of the two photons. This leads a decrease in the photoproduction cross-section. For the same type of dileptons production in the same center-of-mass energy, the production cross-section of photoproduction increases with the charge density of the incoming heavy ion in ultraperipheral heavy-ion collisions, since the higher charge density leads to an increasing of the electromagnetic field intensity. Our calculations show that the large values of the differential cross sections for ultraperipheral collisions can be obtained with the semi-coherent approach at the RHIC and LHC energies.

## V. ACKNOWLEDGEMENTS

This work is supported by Heilongjiang Science Foundation Project under grant No. LH2021A009, National Natural Science Foundation of China under grant No. 12063006, and Special Basic Cooperative Research Programs of Yunnan Provincial Undergraduate Universities Association under grant No. 202101BA070001-144.  the National Natural Science Foundation of China under grantGrant No. 12165010, the Yunnan Province Applied Basic Research Project under grant No. 202101AT070145, the Xingdian Talent Support Project, the Young Top-notch Talent of Kunming, and the Program for Frontier Research Team of Kunming University 2023.